\documentclass[12pt]{iopart}
\usepackage{iopams}  
\usepackage{graphicx}
\usepackage{cases}
\usepackage{url}

\newcommand{\be}{\begin{equation}}
\newcommand{\ee}{\end{equation}}
\newcommand{\bea}{\begin{eqnarray}}
\newcommand{\eea}{\end{eqnarray}}
  
\newcommand{\Pp}{\mathcal{P}}  
\newcommand{\der}[2]{\frac{\partial #1}{\partial #2}}
\newcommand{\D}{\mathrm{d}}  
\newcommand{\Obs}{\mathcal{O}}  
\newcommand{\ObsE}{\Obs_{\mathrm{E}}}  
\newcommand{\Lie}[1]{\pounds_{\bi{#1}}\,}
\newcommand{\LL}{\mathcal{L}}  
\newcommand{\eb}{e}  

\begin{document}

\title{3+1 geodesic equation and images in numerical spacetimes}

\author{F H Vincent$^{1,2}$,
E Gourgoulhon$^{2}$ and J Novak$^{2}$}
\address{$^{1}$ LESIA, Observatoire de Paris, CNRS, Universit\'e Pierre et Marie Curie, Universit\'e Paris Diderot, 5 place Jules Janssen, 92190 Meudon, France\\
$^{2}$ LUTH, Observatoire de Paris, CNRS, Universit\'e Paris Diderot, 5 place Jules Janssen, 92190 Meudon, France}
\ead{frederic.vincent@obspm.fr, eric.gourgoulhon@obspm.fr, jerome.novak@obspm.fr}

\begin{abstract}
The equations governing null and timelike geodesics are derived within the 3+1 formalism of general relativity. In addition to the particle's position, they encompass an evolution equation for the particle's energy leading to a 3+1 expression of the redshift factor for photons. An important application is the computation of images and spectra in spacetimes arising from numerical relativity, via the ray-tracing technique. This is illustrated here by images of numerically computed stationary neutron stars and dynamical neutron stars collapsing to a black hole.
\end{abstract}



\section{Introduction}

The computation of trajectories of photons or test-mass particles in the Kerr
metric is a topic of major importance in relativistic astrophysics. This notably allows
the investigation of spacetime properties around black holes (see e.g.~\cite{fabian89,karas92,paumard08,hamaus09,dovciak04,StrauVAGP12} and references therein), the aim being to determine the black hole's mass and spin and to test
general relativity (GR). Photons and test-mass particles follow spacetime null and timelike geodesics, respectively. Their motion is thus governed by the so-called \emph{geodesic equation}.

However, within the framework of metric theories, strong tests of GR require to compare the Kerr geometry with geometries generated by alternative models of compact objects. The metric is then generally not known analytically and must be computed numerically. Rotating gravastars and boson stars are examples of such objects. Numerical metrics being almost exclusively computed using the 3+1 formalism of GR (see e.g. \cite{Gourg12}), it is quite useful to derive the geodesic equation within this framework. This is a way to build an optimized ray-tracing algorithm.

In addition to the GR tests around astrophysical black holes, another field of application is the visualization of computer-generated spacetimes, 
resulting from numerical relativity studies of sources of gravitational 
radiation, such as gravitational collapse or coalescing binary compact objects \cite{Alcub08,BaumgS10}. 
Such spacetimes are generally computed with the 3+1 formalism, and this motivates
the design of a ray-tracing algorithm based on a 3+1 geodesic equation.

A ray-tracing code capable of using a 3+1 metric, \texttt{GYOTO}, has recently been developed in our group \cite{vincent11}. This code is written in C++, is open source and can be freely downloaded from \cite{gyoto}. It computes null and timelike geodesics, both in the Kerr metric and in any numerically computed spacetime. The goal of this article is to derive the 3+1 geodesic equation that allows \texttt{GYOTO} to compute images and spectra in numerically generated spacetimes, and to give the first examples of astrophysical interest of this capability. To our knowledge, in previous works, the geodesic equation has only been integrated in numerical spacetimes for the purpose of locating event horizons \cite{HugheKWWST94,LibsoMSSW96,CohenPS09}, but not to form images nor to compute spectra. 

The plan of the article is as follows. Section~\ref{sec:31eq} derives the 3+1 geodesic equation. Section~\ref{sec:red} derives the 3+1 expression of the redshift factor, useful for ray-tracing computations. 
Section~\ref{sec:app} presents the first applications of ray-tracing in numerical spacetimes considering stationary and collapsing neutron star spacetimes. 
Finally, Section~\ref{sec:conc} gives conclusions and perspectives for future works.


\section{3+1 geodesic equation}
\label{sec:31eq}

\subsection{Framework} \label{s:framework}

Let $(\mathcal{M},g_{\alpha\beta})$ be a 4-dimensional spacetime, i.e. a 4-dimensional smooth manifold
$\mathcal{M}$ endowed with a pseudo-Riemannian metric $g_{\alpha\beta}$, of signature 
$(-,+,+,+)$. We denote by $\nabla_\alpha$ the Levi-Civita connection associated with $g_{\alpha\beta}$. 
The 3+1 formalism of GR (see e.g. \cite{Gourg12,Alcub08,York79}) is based on the assumption that $(\mathcal{M},g_{\alpha\beta})$ is globally hyperbolic, so that it can be foliated
by a one-parameter family of spacelike hypersurfaces $(\Sigma_t)_{t\in\mathbb{R}}$. 
Let $n^\alpha$ be the future-directed unit normal to the hypersurface $\Sigma_t$. 
$n^\alpha$ is collinear to the gradient of $t$, the proportionality factor defining the \emph{lapse function} $N$: $n_\alpha = - N \nabla_\alpha t$.
The unit timelike vector $n^\alpha$ is the 4-velocity of the so-called 
\emph{Eulerian observers} $\ObsE$, i.e. the observers whose worldlines are orthogonal to the hypersurfaces $\Sigma_t$. 

Using standard notations, we denote by $\gamma_{\alpha\beta}$ the metric
induced by $g_{\alpha\beta}$ on each hypersurface $\Sigma_t$ \emph{(first fundamental form)}. Since $\Sigma_t$ is assumed to be spacelike, $\gamma_{\alpha\beta}$ is a Riemannian metric (i.e. positive definite). One has
\be
\label{e:def_gij}
  \gamma_{\alpha\beta} = g_{\alpha\beta} + n_\alpha n_\beta 
\ee
and $\gamma^\alpha_{\ \, \beta}$ is the orthogonal projector onto $\Sigma_t$. 
We denote by $D_\alpha$ the Levi-Civita connection associated with the metric $\gamma_{\alpha\beta}$ on $\Sigma_t$. 
The 4-acceleration of an Eulerian observer is
$a^\alpha := n^\mu \nabla_\mu n^\alpha$ and obeys 
\be \label{e:a_DN}
  a^\alpha = D^\alpha \ln N . 
\ee
In particular, $n_\mu a^\mu = 0$, so that $a^\alpha$ is tangent to $\Sigma_t$. 
 
The \emph{extrinsic curvature tensor}, or \emph{second fundamental form}, of the hypersurface
$\Sigma_t$ is defined by 
\be
  K_{\alpha\beta} := - \gamma^\mu_{\ \, \alpha} \gamma^\nu_{\ \, \beta} 
  \nabla_\mu n_\nu . 
\ee
One has $K_{\alpha\mu} n^\mu =0$ as well as the useful relation (see e.g. \cite{Gourg12})
\be \label{e:nab_n}
  \nabla_\beta n_\alpha = - K_{\alpha\beta} - D_\alpha \ln N \, n_\beta . 
\ee

In this article, we shall consider only coordinate systems on $\mathcal{M}$ that are \emph{adapted to the 3+1 foliation} $(\Sigma_t)_{t\in\mathbb{R}}$, i.e. coordinate system $(x^\alpha)$ such that $x^0 = t$. 
The three remaining coordinates\footnote{Latin indices span $\{1,2,3\}$, whereas Greek indices span $\{0,1,2,3\}$.} $(x^i)$
span the hypersurfaces $\Sigma_t$: by construction the vectors 
$\partial/\partial x^i$ are tangent to $\Sigma_t$. The vector $\partial/\partial t$ is transverse to $\Sigma_t$ and its 3+1 decomposition defines the
\emph{shift vector} $\beta^\alpha$:
\be \label{e:decom_dsdt}
  \left( \der{}{t} \right)^\alpha = N n^\alpha + \beta^\alpha, 
\quad\mbox{with}\quad n_\mu \beta^\mu = 0 . 
\ee
The knowledge of the lapse function $N$, the shift vector\footnote{We write
$\beta^i$ when we consider the shift as a tangent vector on the manifold $\Sigma_t$ and $\beta^\alpha$ when we consider it as a vector on $\mathcal{M}$, as in (\ref{e:decom_dsdt}); then $\beta^0=0$. We extend this notation to all tensor fields \emph{tangent to $\Sigma_t$}, in the sense that their contraction with $n_\alpha$
or $n^\alpha$ vanishes (for instance $\gamma_{\alpha\beta}$ or the vector $V^\alpha$ introduced below).} $\beta^i$ and
the induced metric $\gamma_{ij}$ is sufficient to reconstruct the 
spacetime metric $g_{\alpha\beta}$ according to 
\be
  g_{\mu\nu} \, \D x^\mu \, \D x^\nu = 
  - N^2 \D t^2 + \gamma_{ij} (\D x^i + \beta^i \D t)
		(\D x^j + \beta^j \D t) . 
\ee 
In the 3+1 formalism, the solution of the Einstein equation is obtained by solving a Cauchy problem (with constraints) for $(\gamma_{ij},K_{ij})$: some initial 
values being given on a hypersurface $\Sigma_0$, obeying the constraint equations, the fundamental forms $(\gamma_{ij},K_{ij})$ are evolved in time to construct the whole 
spacetime~\cite{Gourg12,Alcub08,BaumgS10}. In this formulation, the lapse
and shift are not dynamical variables; their role is to set the coordinates. 

\subsection{Geodesic equation in 3+1 covariant form} \label{s:geod_congruence}

Let us consider a particle $\Pp$ of 4-momentum $p_\alpha$. $\Pp$ can either be
a photon, in which case $p_\mu p^\mu = 0$, or a massive particle, of mass 
$m = \sqrt{-p_\mu p^\mu}$. If $\Pp$ is subject only to the gravitational field, its worldline 
$\LL$ is a either a null (photon), or a timelike (massive particle) geodesic of $(\mathcal{M},g_{\alpha\beta})$. The 4-momentum then obeys
\be \label{e:p_geod}
  p^\mu \nabla_\mu p^\alpha = 0 . 
\ee
This is the geodesic equation in covariant 4-dimensional form. To write it in 3+1 form, we start by performing an orthogonal decomposition of $p^\alpha$
according to 
\be \label{e:decomp_p}
  p^\alpha = E (n^\alpha + V^\alpha), 
\quad\mbox{with}\quad n_\mu V^\mu = 0 . 
\ee
The scalar $E$ is nothing but the energy of $\Pp$ as measured by the Eulerian observer $\ObsE$; indeed, $n^\alpha$ is the 4-velocity of $\ObsE$, and then 
(\ref{e:decomp_p}) implies $E=-p_\mu n^\mu$. 
The vector $V^\alpha$ is by construction tangent to $\Sigma_t$ and
coincides with the 3-velocity of $\Pp$ as measured by $\ObsE$. 
To show it, we notice that the orthogonal projection $P^\alpha := \gamma^\alpha_{\ \mu} p^\mu$
of $p^\alpha$ on $\ObsE$'s rest space is the linear 3-momentum of $\Pp$ with respect to $\ObsE$ and (\ref{e:decomp_p}) implies $P^i = E V^i$, which 
is exactly the relation between the 3-momentum and the 3-velocity of a 
particle (massive or not) of energy $E$. 
For a photon, the property $p_\mu p^\mu = 0$, together with 
$n_\mu n^\mu = -1$ and (\ref{e:decomp_p}), imply $V_\mu V^\mu = V_i V^i = 1$, i.e. with respect to
$\ObsE$, the photon travels at the speed of light (as it should!). 
For a massive particle, one has instead
$p_\mu p^\mu < 0$. The assumption $E>0$, (\ref{e:decomp_p}) leads then to 
$V_i V^i < 1$: $\Pp$ cannot reach the speed of light. 

In the remaining part of this section, we do not consider a single geodesic, but a full congruence of them. 
This means that $p^\alpha$, $E$ and $V^\alpha$ are fields defined on the spacetime. Accordingly, we may consider their derivatives in any direction and not only along a geodesic as in (\ref{e:p_geod}). 
Let us then rewrite (\ref{e:p_geod}) by substituting (\ref{e:decomp_p}) for $p^\alpha$ and making use of 
(\ref{e:a_DN}) and (\ref{e:nab_n}) to express $n^\mu \nabla_\mu n^\alpha$ 
and $\nabla_\mu n^\alpha$; we get
\be \label{e:geod_3p1_prov}
  \fl (n^\mu + V^\mu) \nabla_\mu E \, (n^\alpha + V^\alpha)
  + E \left( D^\alpha \ln N +  n^\mu \nabla_\mu V^\alpha - K^\alpha_{\ \,\mu} V^\mu 
  +  V^\mu  \nabla_\mu V^\alpha \right) = 0 .
\ee
Now, in the 3+1 formalism, the natural evolution operator is the Lie derivative $\Lie{m}$ along the vector field
$m^\alpha := N n^\alpha$, for it preserves the property of being tangent 
to $\Sigma_t$ \cite{Gourg12}. 
Therefore we write
\bea
  n^\mu \nabla_\mu V^\alpha & = & N^{-1} m^\mu \nabla_\mu V^\alpha 
  = N^{-1} \left[ \Lie{m} V^\alpha + V^\mu\nabla_\mu (N n^\alpha) \right] \nonumber \\
 & = &N^{-1} \Lie{m} V^\alpha - K^\alpha_{\ \, \mu} V^\mu
  + V^\mu D_\mu \ln N \, n^\alpha . \label{e:n_nab_V}
\eea
Similarly, since $E$ is a scalar field, 
\be \label{e:n_nab_E}
  n^\mu \nabla_\mu E = N^{-1}  \Lie{m} E . 
\ee
Also, the 4-dimensional and 3-dimensional covariant derivatives of $V^\alpha$ are related by
\be
  D_\beta V^\alpha = \gamma^\alpha_{\ \, \mu} \gamma^\nu_{\ \, \beta} \nabla_\nu V^\mu, 
\ee
from which we deduce the identity
\be \label{e:V_nab_V}
  V^\mu \nabla_\mu V^\alpha = V^\mu D_\mu V^\alpha - K_{\mu\nu} V^\mu V^\nu \, n^\alpha . 
\ee
Note that, as in many places in the article, we have used the property $n_\mu V^\mu = 0$ along with
expression (\ref{e:nab_n}) for $\nabla_\beta n^\alpha$. 

Inserting (\ref{e:n_nab_V}), (\ref{e:n_nab_E}) and (\ref{e:V_nab_V}) into (\ref{e:geod_3p1_prov}), we get, after
division by $E$,
\bea
   \fl  N^{-1} \Lie{m} V^\alpha  + V^\mu D_\mu V^\alpha - 2 K^\alpha_{\ \, \mu} V^\mu
  + E^{-1} ( N^{-1}  \Lie{m} E + V^\mu D_\mu E ) V^\alpha  + D^\alpha \ln N
    \nonumber \\
    + \left[ V^\mu D_\mu \ln N - K_{\mu\nu} V^\mu V^\nu  
  + E^{-1} ( N^{-1}  \Lie{m} E + V^\mu D_\mu E ) \right] n^\alpha = 0 . \label{e:geod_3p1_prov2}
\eea
Let us notice that the first line of this equation contains only terms 
tangent to $\Sigma_t$, whereas the term in the second line is manifestly
parallel to $n^\alpha$. Hence the projections of (\ref{e:geod_3p1_prov2}) along $n^\alpha$ and onto $\Sigma_t$ give respectively
\bea
  \fl  N^{-1}  \Lie{m} E + V^\mu D_\mu E  +  E \left( 
   V^\mu D_\mu \ln N  - K_{\mu\nu} V^\mu V^\nu  \right) = 0 \label{e:evol_E_m}\\
  \fl N^{-1} \Lie{m} V^\alpha  + V^\mu D_\mu V^\alpha - 2 K^\alpha_{\ \, \mu} V^\mu
  + E^{-1} ( N^{-1}  \Lie{m} E + V^\mu D_\mu E ) V^\alpha  + D^\alpha \ln N = 0 . 
  \label{e:evol_V_m}
\eea
These two equations involve only quantities intrinsic to $\Sigma_t$. 
We may then write them in a 3-dimensional form, using (\ref{e:decom_dsdt}) to express the
Lie derivative along $m^\alpha$ :
\be \label{e:Lie_m_t_beta}
  \Lie{m} = \der{}{t} - \Lie{\beta} .
\ee
If, in addition, we substitute (\ref{e:evol_E_m}) for $\Lie{m} E$ in (\ref{e:evol_V_m}),
we get
\bea
  \fl \frac{1}{N} \left( \der{}{t} - \Lie{\beta} \right) E 
  + V^j D_j E + E (V^j D_j \ln N - K_{jk} V^j V^k ) = 0 \label{e:dEdt_cov} \\
  \fl \frac{1}{N} \left( \der{}{t} - \Lie{\beta} \right) V^i 
  + V^j D_j V^i - 2 K^i_{\ \, j} V^j + V^i (K_{jk} V^j V^k  - V^j D_j \ln N) \nonumber \\
  + D^i \ln N = 0 . \label{e:dVdt_cov}
\eea
This system constitutes the 3+1 geodesic equation in covariant form for a congruence of geodesics. 

\subsection{3+1 geodesic equation for a single geodesic} \label{s:3p1_geod_coord}

Let us consider a specific member $\LL$ of the geodesic congruence, representing the worldline
of a given particle $\Pp$. In a coordinate system $(x^\alpha) = (t,x^i)$
adapted to the 3+1 foliation $(\Sigma_t)_{t\in\mathbb{R}}$ (cf. Section~\ref{s:framework}),
the equation of $\LL$ can be written as 
\be \label{e:X_t}
  x^i = X^i(t) ,
\ee
where the $X^i$'s are three smooth functions $\mathbb{R}\rightarrow\mathbb{R}$. 
This is nothing but a parametrization of $\LL$ by the coordinate time $t$.
Note that, a priori, $t$ is not an affine parameter along $\LL$.

\begin{figure}
\centering
	\includegraphics[width=0.7\textwidth]{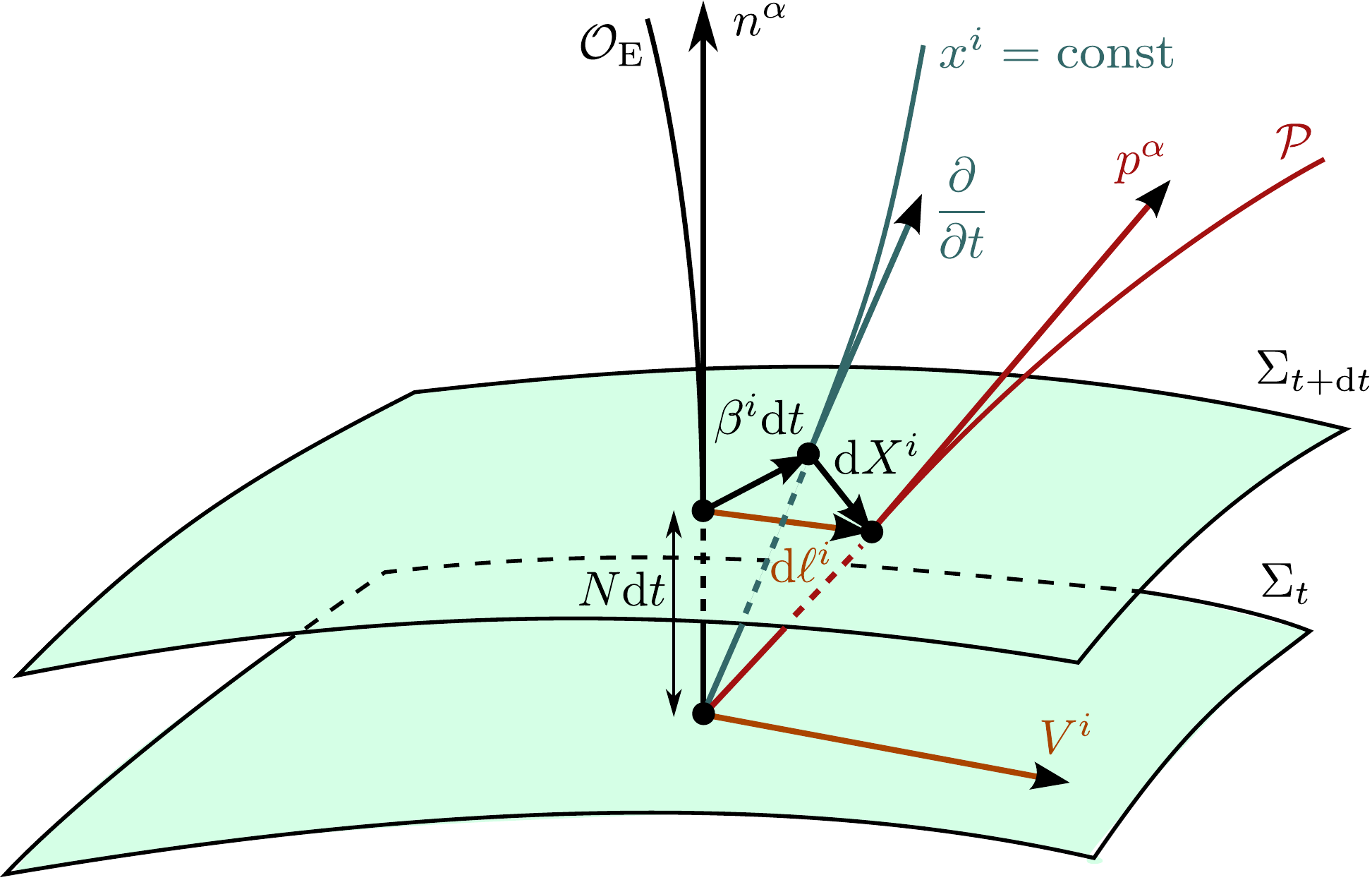}
	\caption{Relation between the 3-velocity of $\Pp$ relative to the Eulerian observer $V^i=\D\ell^i/(N\D t)$, the coordinate displacement $\D X^i/\D t$ and the shift vector $\beta^i$.}
	\label{f:velocity_shift}
\end{figure}

By definition, the velocity of $\Pp$ with respect to the Eulerian observer $\ObsE$ is
\be
  V^i = \frac{\D \ell^i}{\D\tau_{\mathrm{E}}}, 
\ee
where $\D\ell^i$ is the displacement vector of $\Pp$'s worldline with respect to 
the $\ObsE$'s one between $t$ and $t+\D t$, and $\D\tau_{\mathrm{E}}$ is the increment of
$\ObsE$'s proper time between $t$ and $t+\D t$. 
The origin of the $(x^i)$ coordinates being ``shifted'' by the amount 
$\beta^i \, \D t$ with respect to $\ObsE$'s worldline, we have
(cf. Figure~\ref{f:velocity_shift})
\be
  \D\tau_{\mathrm{E}} = N\, \D t \quad\mbox{and}\quad \D\ell^i = \beta^i \D t + \D X^i . 
\ee
Hence
\be \label{e:V_dXdt}
  V^i = \frac{1}{N} \left( {\dot X}^i + \beta^i \right) ,\quad\mbox{with}\quad
  {\dot X}^i := \frac{\D X^i}{\D t} . 
\ee

The variation of $E$ along $\LL$ associated with the parametrization by $t$ is
\be \label{e:dedt_lagangien}
  \frac{\D E}{\D t} = \der{E}{t} + {\dot X}^j \partial_j E, 
\ee
where $\partial_j := \partial / \partial x^j$.
Also, since $E$
is a scalar field, 
$\Lie{\beta} E = \beta^j \partial_j E$ and $V^j D_j E = V^j \partial_j E$. 
Thanks to (\ref{e:V_dXdt}),
(\ref{e:dEdt_cov}) can be then written as
\be \label{e:dEdt}
  \frac{\D E}{\D t} = E \left( N K_{jk} V^j V^k - V^j \partial_j N \right) . 
\ee
This evolution equation for the particle energy relative to the Eulerian observer is
equivalent to equation (6) of Ref.~\cite{merlin11}. The latter is actually an evolution equation in term
of some affine parameter $\lambda$ along the geodesic, i.e. a parameter 
whose associated tangent vector is (up to some constant factor) the particle's 4-momentum: 
$p^\alpha = \D x^\alpha / \D \lambda$. Then,
$n_\mu p^\mu \D \lambda = n_\mu \D x^\mu$, which results in $- E \D \lambda = - N \D t$. Hence the 
relation between the two parametrizations of $\LL$:
\be \label{e:dlambdadt}
    \frac{\D \lambda}{\D t} = \frac{N}{E} . 
\ee
Taking into account (\ref{e:dlambdadt}), we check that (\ref{e:dEdt}) is indeed equivalent to 
(6) of Ref.~\cite{merlin11}.

Let us now consider equation (\ref{e:dVdt_cov}); we may express the Lie and covariant derivatives 
in terms of partial derivatives:
\[
  \fl \frac{1}{N} \left( \der{}{t} - \Lie{\beta} \right) V^i 
  + V^j D_j V^i = \frac{1}{N} \left( \der{V^i}{t} 
  - \beta^j \partial_j V^i + V^j\partial_j \beta^i \right)
  + V^j \partial_j V^i + {}^3\Gamma^i_{jk} V^j V^k , 
\]
where the ${}^3\Gamma^i_{jk}$'s are the Christoffel symbols of the metric $\gamma_{ij}$
in $\Sigma_t$. Then, 
by means of (\ref{e:V_dXdt}) and the analog of (\ref{e:dedt_lagangien}) for $V^i$:
\be 
  \frac{\D V^i}{\D t} = \der{V^i}{t} + {\dot X}^j \partial_j V^i, 
\ee
we transform (\ref{e:dVdt_cov}) into
\[
    \frac{\D V^i}{\D t} = N V^j\left[ V^i\left( \partial_j\ln N - K_{jk}  V^k \right)
  + 2 K^i_{\ \, j}  - {}^3\Gamma^i_{jk} V^k \right]  - \gamma^{ij}\partial_j N - V^j \partial_j \beta^i.
\]
Let us supplement this equation by the evolution equation for $X^i$ deduced from (\ref{e:V_dXdt}), to form the system 
\begin{subnumcases}{\label{e:systXV}}
  \frac{\D X^i}{\D t} = & $ \!\!\!\!\!\! N V^i - \beta^i$ \label{e:dXdt} \\[1ex]
  \frac{\D V^i}{\D t} = &   $\!\!\!\!\!\! N V^j \left[ V^i\left( \partial_j\ln N - K_{jk} V^k \right)
  + 2 K^i_{\ \, j} - {}^3\Gamma^i_{jk} V^k \right] - \gamma^{ij}\partial_j N - V^j \partial_j \beta^i$ . \label{e:dVdt}
\end{subnumcases}

Given the spacetime metric in 3+1 form, and hence the terms $N$, $\beta^i$, $\gamma^{ij}$, 
${}^3\Gamma^i_{jk}$ and $K_{ij}$, (\ref{e:systXV}) constitutes a system of six first order ordinary differential equations, that it is sufficient to integrate with respect to $t$ from initial data $(X^i(0),V^i(0))$ to get the geodesic worldline of $\Pp$ in the form (\ref{e:X_t}). 

Note that, contrary to (\ref{e:dEdt_cov})-(\ref{e:dVdt_cov}), the energy equation (\ref{e:dEdt})
and the system (\ref{e:systXV}) are meaningful for a single geodesic: they involve only the derivatives 
$\D E/\D t$, $\D X^i/\D t$ and $\D V^i/\D t$, which are derivatives \emph{along} the geodesic 
(and not transverse to it). To strengthen this point, 
we present in \ref{s:derive_single} an alternative derivation
of (\ref{e:dEdt}) and (\ref{e:systXV}), which does not rely on the assumption that $\LL$ belongs 
to some geodesic congruence. 

A 3+1 form of the geodesic equation has already been derived by Hughes et al. \cite{HugheKWWST94}
(cf. also Section 7.2 of the textbook \cite{BaumgS10}). However, it differs from the present one by the following
features: (i) it involves the components $p_i$ of the 4-momentum instead of $V^i$, 
(ii) the evolution parameter is the affine parameter $\lambda$ and not the coordinate time $t$,
(iii) it is valid only for massless particles. Moreover, in Ref.~\cite{HugheKWWST94} no evolution equation
for the particle's energy, equivalent to our equation~(\ref{e:dEdt}) is provided. Note also that 
the 3+1 geodesic system of \cite{HugheKWWST94} is applied to the determination of the event horizon, not to the formation of images. 

In \ref{s:second_order}, we combine (\ref{e:dXdt}) and (\ref{e:dVdt}) into a single second order equation for $X^i(t)$. We recover in this way the standard 4-dimensional geodesic equation. 
However, this equation is less convenient for numerical integration for it involves time derivatives of the lapse and shift, contrary to the system (\ref{e:systXV}). 


\section{Redshift factor}
\label{sec:red}

\subsection{General formula}

The integration of (\ref{e:dEdt}) forward in $t$ gives the energy of 
the particle $\Pp$ with respect to the Eulerian observer at any point.
Let us consider the case in which $\Pp$ is a photon emitted at some event $A$ by 
an observer $\Obs_{\rm em}$ (the ``emitter'') and received at some event $B$ by an observer $\Obs_{\rm rec}$ (the ``receiver''). Note that these observers are not necessarily Eulerian observers. 
In this way, the problem considered here generalizes that of Ref.~\cite{merlin11}, which was
limited to Eulerian observers. 
The redshift factor $z$ is defined by
\be
  1+z = \frac{\nu_{\rm em}}{\nu_{\rm rec}} , 
\ee
where $\nu_{\rm em}$ (resp. $\nu_{\rm rec}$) is the photon frequency measured by 
$\Obs_{\rm em}$ (resp. $\Obs_{\rm rec}$).
The frequency being related to the energy by the Planck-Einstein formula $\varepsilon = h \nu$, 
the above relation can be written
\be \label{e:redshift_prov}
  1+z = \frac{\varepsilon_{\rm em}}{\varepsilon_{\rm rec}} 
  = \frac{\left. p_\mu\right| _A u^\mu_{\rm em}}{\left. p_\mu \right| _B u^\mu_{\rm rec}}, 
\ee
where $\varepsilon_{\rm em} = - \left. p_\mu\right| _A u^\mu_{\rm em}$ (resp. $\varepsilon_{\rm rec} = - \left. p_\mu \right| _B u^\mu_{\rm rec}$) is the photon energy with respect to 
$\Obs_{\rm em}$ (resp. $\Obs_{\rm rec}$) and $u^\alpha_{\rm em}$ (resp. $u^\alpha_{\rm rec}$)
is the 4-velocity of $\Obs_{\rm em}$ at $A$ (resp. of $\Obs_{\rm rec}$ at $B$). 
Let us perform the 3+1 decomposition of these 4-velocities: 
\bea
  u^\alpha_{\rm em} = \Gamma_{\rm em} (n^\alpha + U^\alpha_{\rm em})
  \quad\mbox{with}\quad
  n_\mu U^\mu_{\rm em} = 0 \label{e:decomp_uem} \\
  u^\alpha_{\rm rec} = \Gamma_{\rm rec} (n^\alpha + U^\alpha_{\rm rec}) 
  \quad\mbox{with}\quad
  n_\mu U^\mu_{\rm rec} = 0.
\eea
$\Gamma_{\rm em}$ (resp. $\Gamma_{\rm rec}$) is then the Lorentz factor of $\Obs_{\rm em}$ 
(resp. $\Obs_{\rm rec}$) with respect to the Eulerian observer
$\ObsE$ and $U^\alpha_{\rm em}$ (resp. $U^\alpha_{\rm rec}$) the 3-velocity of $\Obs_{\rm em}$ 
(resp. $\Obs_{\rm rec}$) with respect to
$\ObsE$. From the normalization relation $u_\mu u^\mu =-1$, we get 
\be \label{e:Lorentz_fact}
  \Gamma_{\rm em} = \left( 1 - \gamma_{ij} U_{\rm em}^i U_{\rm em}^j \right) ^{-1/2} 
  \quad\mbox{and}\quad
  \Gamma_{\rm rec} = \left( 1 - \gamma_{ij} U_{\rm rec}^i U_{\rm rec}^j \right) ^{-1/2} .
\ee
Combining (\ref{e:decomp_p}) and (\ref{e:decomp_uem}) yields
\be \label{e:eps_em_E_A}
  \varepsilon_{\rm em} = - \left. p_\mu\right| _A u^\mu_{\rm em} = 
  \left. E \right| _A \Gamma_{\rm em} \left( 1 - \gamma_{ij} \left. V^i \right| _A U_{\rm em}^j 
  \right) .
\ee
A similar relation holds for $\varepsilon_{\rm rec}$, so that (\ref{e:redshift_prov}) becomes, once (\ref{e:Lorentz_fact}) is taken into
account, 
\be \label{e:redshift}
  1 + z = \frac{ \left. E \right| _A }{ \left. E \right| _B } \; 
  \frac{ 1 - \gamma_{ij} \left. V^i \right| _A U_{\rm em}^j }{ 1 - \gamma_{ij} \left. V^i \right| _B U_{\rm rec}^j} \; 
  \left( \frac{ 1 - \gamma_{ij} U_{\rm rec}^i U_{\rm rec}^j}{1 - \gamma_{ij} U_{\rm em}^i U_{\rm em}^j} \right) ^{1/2}  . 
\ee
The procedure to compute the redshift factor is then as follows. 
Given some initial data\footnote{In a ray-tracing code, the integration is
usually performed backward, i.e. from $B$ to $A$. Accordingly, the roles of $A$ and $B$ have to be swapped in the following discussion.} $\left. E \right| _A = E(t=t_A)$, 
$\left. V^i \right| _A = V^i(t=t_A)$ and $X^i(t=t_A) = x^i_A$, where $(x^i_A)$
are the coordinates of the event $A$ on the hypersurface $\Sigma_{t_A}$, 
one integrates the system formed by (\ref{e:dEdt}) and
(\ref{e:systXV}) to get
$\left. E \right| _B = E(t=t_B)$ and $\left. V^i \right| _B = V^i(t=t_B)$.
Then, one uses (\ref{e:redshift}) to evaluate $z$. 
Note that since (\ref{e:dEdt}) is a homogeneous equation in $E$ and only the ratio
$\left. E \right| _A / \left. E \right| _B$ is involved in the
expression of $z$, the initial value
$\left. E \right| _A$ can be chosen arbitrarily. Of course, if one wants to
manipulate some physically
relevant value of $E$, one may deduce $\left. E \right| _A$ from the photon energy with respect to the
emitter, $\varepsilon_{\rm em}$, via (\ref{e:eps_em_E_A}). 

\subsection{Limiting cases}

As a check of formula (\ref{e:redshift}), let us consider the special case of an inertial observer in Minkowski spacetime receiving a photon emitted by a moving source. Choosing for 
$(\Sigma_t)_{t\in\mathbb{R}}$ the time foliation associated with that inertial observer, we have
$\ObsE = \Obs_{\rm rec}$, so that $U^i_{\rm rec} = 0$. Moreover, $N=1$ and $K_{ij} = 0$, so that
(\ref{e:dEdt}) reduces to $\D E / \D t = 0$, implying $\left. E \right| _B = \left. E \right| _A$. 
Accordingly (\ref{e:redshift}) reduces to 
\be \label{e:redshift_special}
  1 + z = \frac{ 1 - f_{ij} \left. V^i \right| _A U_{\rm em}^j }{\sqrt{ 1 - f_{ij} U_{\rm em}^i U_{\rm em}^j}} , 
\ee
where $f_{ij}$ is the flat metric. We recover the special relativistic formula for the Doppler effect. 
In particular, if the photon travels in the same direction (up to a sign) as the emitter, we have 
$f_{ij} \left. V^i \right| _A U_{\rm em}^j = U$ and 
$f_{ij} U_{\rm em}^i U_{\rm em}^j = U^2$ with $U := \pm \sqrt{f_{ij} U_{\rm em}^i U_{\rm em}^j}$,
with a $+$ (resp. $-$) sign if the emitter is approaching to (resp. receding from) 
the receiver, so that (\ref{e:redshift_special}) gives the well-known formula
\be
  1 + z = \sqrt{ \frac{1 - U}{1+U} } .
\ee

Another check of formula (\ref{e:redshift}) and equation (\ref{e:dEdt}) is provided 
by the propagation of a photon between two static observers in Schwarzschild spacetime. Choosing $(\Sigma_t)$ to be the standard foliation 
associated with Schwarzschild time coordinate $t$, the static observers coincide with Eulerian observers, so that we have $U_{\rm em}^i = 0$ and 
$U_{\rm rec}^i=0$. Accordingly, (\ref{e:redshift}) reduces to 
\be \label{e:z_Schwarz_prov}
  1 + z = \frac{ \left. E \right| _A }{ \left. E \right| _B } . 
\ee
On the other side, (\ref{e:dEdt}) reduces to 
\be
  \frac{\D E}{\D t} = - \frac{E}{N} ({\dot X}^j \partial_j N), 
\ee
because $K_{ij} = 0$ for the foliation $(\Sigma_t)$ and $\beta^i = 0$
in standard Schwarzschild coordinates $(t,r,\theta,\varphi)$. 
Since $\partial N / \partial t = 0$, we may rewrite the above equation as
\be
  \frac{\D E}{\D t} = - \frac{E}{N}\frac{\D N}{\D t} ,
\ee
from which we deduce immediately
\be
  \frac{\D}{\D t} (EN) = 0 . 
\ee
Hence $E = \mathrm{const} / N$ and (\ref{e:z_Schwarz_prov}) becomes
$1+z = \left. N \right| _B / \left. N \right| _A$. Using the value of the lapse in terms of the mass parameter $M$ of
Schwarzschild metric and the Schwarzschild coordinate $r$, $N = \sqrt{1- 2M / r}$, we get
\be
  1 + z = \sqrt{ \frac{1- 2M / r_B}{1 - 2M / r_A} } . 
\ee
We recognize the classical formula for the gravitational redshift (Einstein effect). 
In particular, in the ``Sirius B configuration'' ($r_B\rightarrow +\infty$ and 
$M/r_A \ll 1$), we get $z \simeq M / r_A$, as it should be. 


\section{Applications}
\label{sec:app}

\subsection{Implementation in the \texttt{GYOTO} code}

The 3+1 geodesic equations (\ref{e:dEdt}) and (\ref{e:systXV}) have been implemented in the ray-tracing
code \texttt{GYOTO} \cite{vincent11,gyoto}. The integration in $t$ is performed by means of a fourth-order
Runge-Kutta algorithm. 
The 3+1 fields $\left(N, \beta^i, \gamma_{ij}, K_{ij} \right)$ have to
be provided by an external code. An example of using the 3+1 fields from a numerical relativity spectral code is given in Section 3 of \cite{vincent11}, where it is shown how to get the values of $\left(N, \beta^i, \gamma_{ij}, K_{ij} \right)$
at each point of the geodesic from the outputs of the spectral code. 

The derivatives with respect to some affine parameter $\lambda$ along the geodesic, $(\D t/\D\lambda,\D r/\D\lambda,\D\theta/\D\lambda,\D\varphi/\D\lambda)$, can be derived from the 3+1 derivatives
$(\D r/\D t,\D\theta/\D t,\D\varphi/\D t)$ provided that one knows the value of $\D t/\D\lambda$.
The latter is deduced from the value of $E$ resulting from the integration of (\ref{e:dEdt}), 
by noticing that $E = -n_{\mu} p^{\mu} = N\,p^t$ and 
$p^t = m u^t = m \, \D t / \D\lambda$ for a massive 
particle ($\lambda$ is then the particle's proper time) and $p^t = \D t / \D \lambda$ (up to some constant
rescaling of $\lambda$) for a photon. 

\begin{figure}
\centering
	\includegraphics[width=0.7\textwidth]{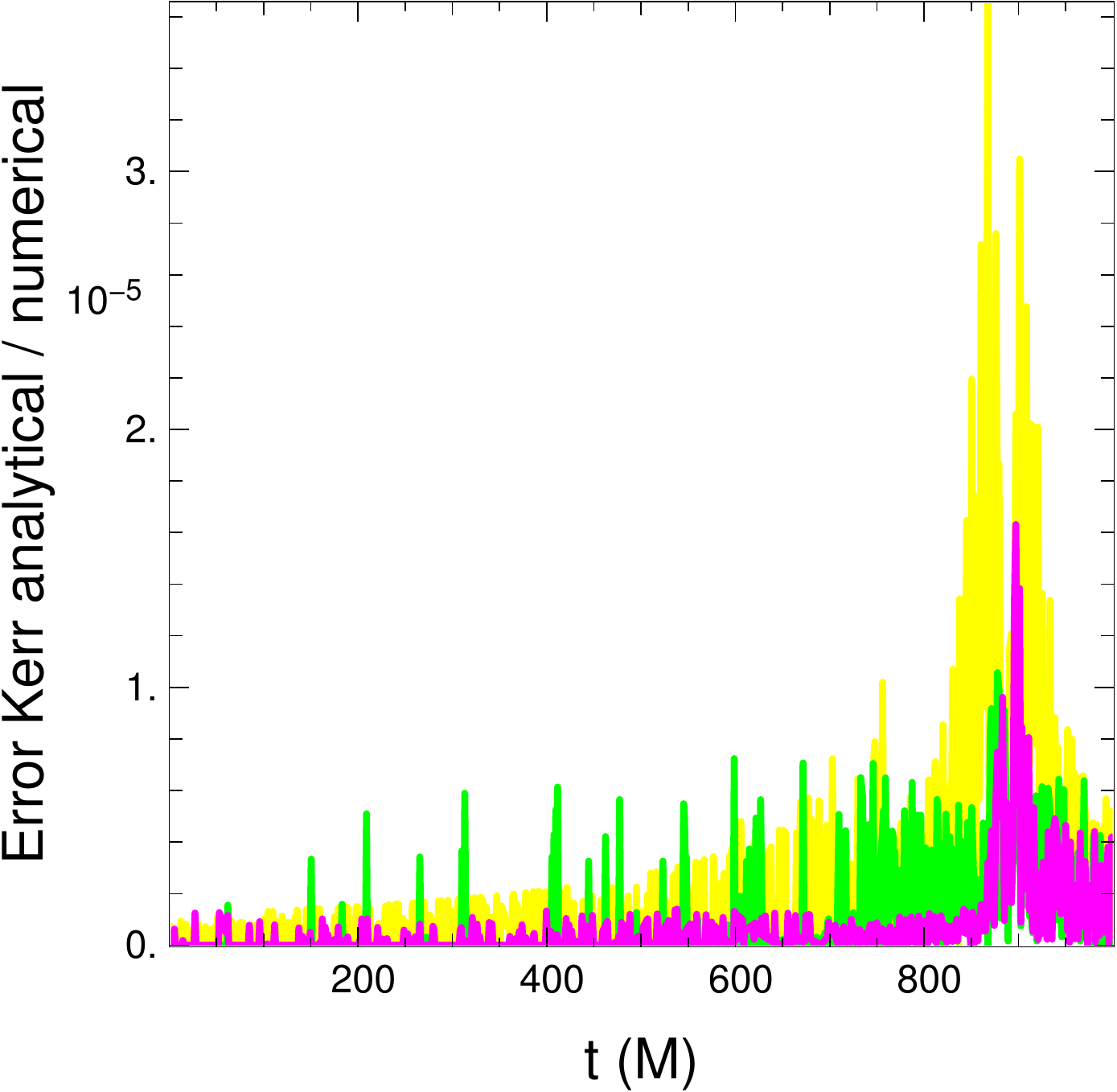}
	\caption{Relative error, in units of $10^{-5}$, on a null geodesic integrated by \texttt{GYOTO} in a numerically computed Kerr metric with spin parameter $a = 0.5\,M$, when compared to the standard integration using the analytical expression of the Kerr metric, for the Boyer-Lindquist coordinates $r$ (yellow), $\theta$ (green) and $\varphi$ (magenta) as a function of time coordinate $t$. The geodesic is integrated backward in coordinate time from $t=1000\,M, r=100\,M$ until $t=0, r=865\,M$. It reaches the smallest distance ($r=4.3\,M$) from the black hole around $t=900\,M$: this is where the error is the largest.}
	\label{f:errkerr}
\end{figure}

\subsection{Numerical tests}

A preliminary test of the 3+1 computation of geodesics in a numerical spacetime has been provided in Figure~8 of \cite{vincent11},
where a timelike geodesic computed by the 3+1 method was compared
to that computed by integrating the standard 4-dimensional geodesic equation 
[equation~(\ref{e:geod_4D}) below]. The spacetime was that of a rapidly rotating relativistic star, numerically generated by means of the \texttt{LORENE/nrotstar} code \cite{gourgoulhon10b,lorene}.

We present here a more detailed test, regarding a null geodesic around a Kerr black hole, with 
a spin parameter $a=0.5\,M$ ($M$ being the black hole mass). The Kerr spacetime is described in 
Boyer-Lindquist coordinates $(t,r,\theta,\varphi)$ and its 3+1 ``numerical'' version has been 
prepared on a spectral grid via a code using the \texttt{LORENE} library \cite{lorene}.
A test null geodesic, that comes close to the event horizon and therefore subject to strong-field effects, 
has been integrated by \texttt{GYOTO} via two methods: (i) integration of the 3+1 geodesic equations in the numerically generated Kerr spacetime, and (ii) standard 4-dimensional integration using the analytical expression of the Kerr metric in Boyer-Lindquist coordinates \cite{vincent11}.

Figure~\ref{f:errkerr} shows the resulting relative difference between the spatial coordinates $(r(t),\theta(t),\varphi(t))$ obtained by the two methods. The maximum relative error, occurring when the geodesic comes at the closest distance to the black hole, is of a few $10^{-5}$, which is very satisfactory. 

\subsection{Images of a stationary rotating neutron star}

Using the 3+1 geodesic equations implemented in \texttt{GYOTO}, we have computed the image perceived by a distant observer of a rapidly rotating neutron star in a spacetime computed by the \texttt{LORENE/nrotstar} numerical code~\cite{gourgoulhon10b,lorene}. 
The neutron star model is built upon an equation of state derived by Akmal, Pandharipande and Ravenhall
\cite{akmal98} and is described in detail in Section~3.5.3 of \cite{gourgoulhon10b}.
The mass of the star in $M=1.4 M_\odot$ and it is chosen to be either non-rotating or rotating 
at the frequency of $716$~Hz (the largest observed frequency \cite{hessels06}). 

\begin{figure}
\centering
	\includegraphics[width=0.4\textwidth]{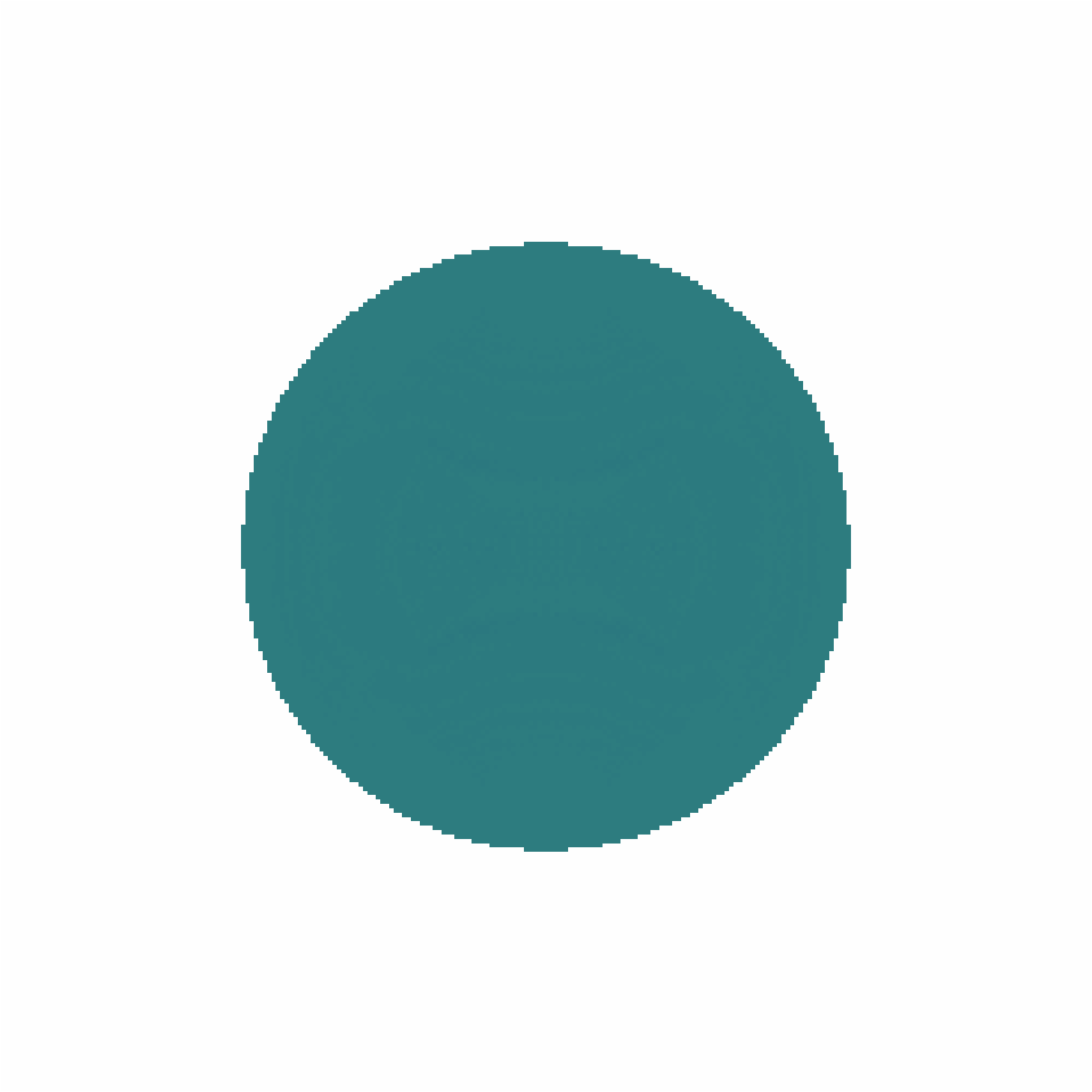}
	\includegraphics[width=0.4\textwidth]{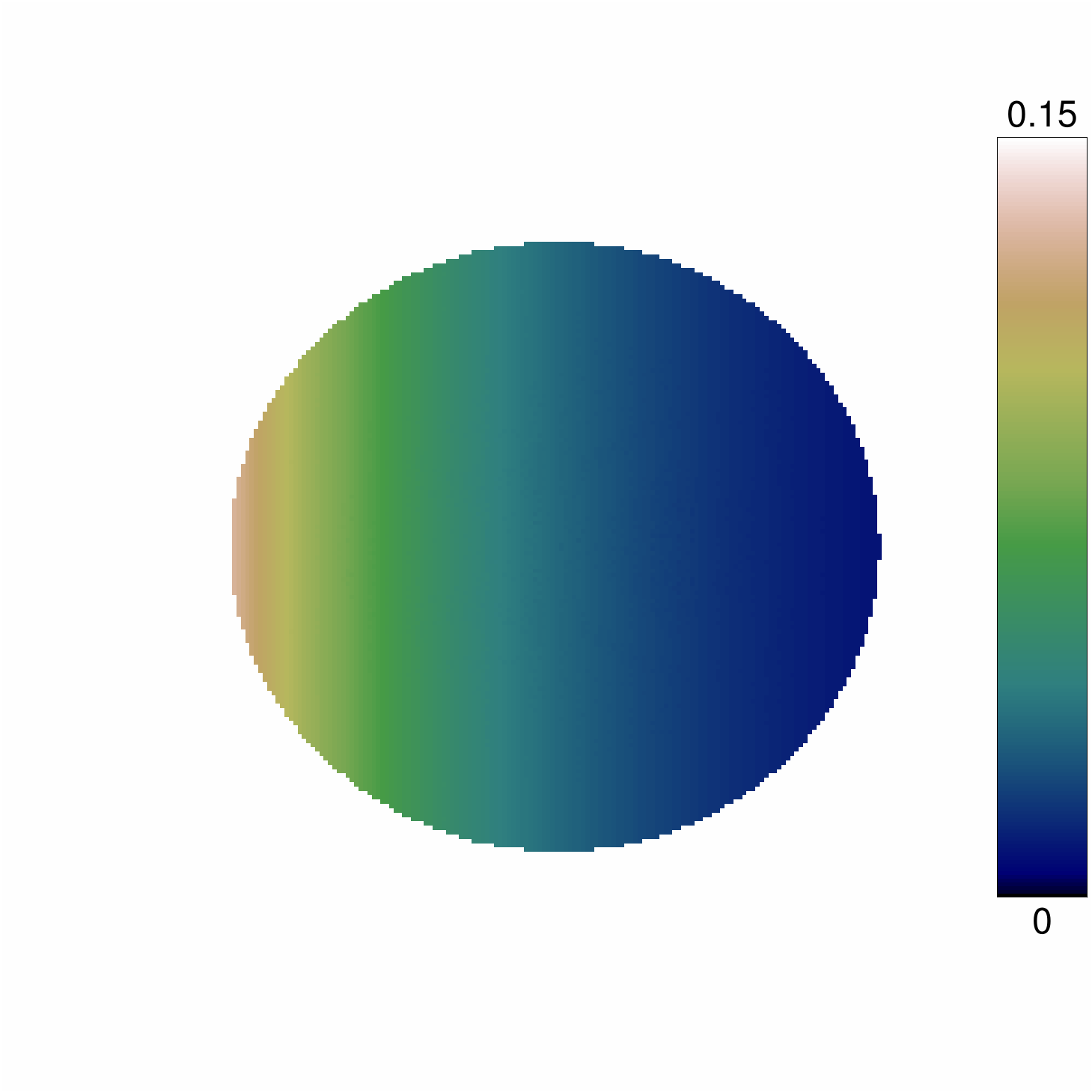}
	\caption{Images (i.e. map of specific intensity) of a non-rotating (left) and $716$~Hz rotating (right) stationary neutron star, with an optically thick surface emitting black body radiation at $10^{6}$~K. The color bar is common to the two panels and is given in SI units, $\mathrm{W\,m^{-2}\,ster^{-1}\,Hz^{-1}}$. The frequency of the photons in the observer's frame is chosen to be $10^{17}$~Hz, close to the maximum of the Planck function at $10^{6}$~K.}
	\label{f:statNS}
\end{figure}

Figure~\ref{f:statNS} shows the image of these two models of neutron stars, assuming the surface of the star is optically thick and emits as a black body at $10^{6}$~K. The effect of relativistic beaming, due to the star's rotation, appears clearly on the right panel.
Moreover, the rotating star is oblate, the ratio of its apparent polar radius to its apparent equatorial radius is 94\%. This effect is a known consequence of its rotation. 

Since the spacetime is stationary and axisymmetric, two quantities must be conserved along 
each geodesic: the components $p_t$ and $p_\varphi$ of the 4-momentum. 
A third constant of motion is the squared norm of the 4-momentum :
$p_\mu p^\mu = 0$ (null geodesic). The constancy of these three quantities is not imposed in the code. 
We therefore monitor them along the geodesics in order to check that the integration is 
performed correctly. In the present case, the squared norm of the photon $p_\mu p^\mu$ stays below a few $10^{-5}$, the maximum relative error on $p_t$ is a few $10^{-6}$ and the maximum relative error on $p_\varphi$ is a few $10^{-4}$.

Let us end this section by a remark that concerns also the next section. Considering a neutron star at a distance of $1$~kpc with a size of $10$~km, its apparent size disregarding any relativistic effect on the photon's trajectory would be less than 
$10^{-10}$ arcsecond. This is of course far beyond the resolution of any current or near future instrument. The present and next sections must therefore not be read as describing possible observational tests.

\subsection{Images of the collapse of a neutron star to a black hole}

To illustrate the 3+1 geodesic computation in dynamical spacetimes, we consider the astrophysical scenario of an unstable non-rotating neutron star collapsing to a black hole. This scenario is numerically modeled using the \texttt{CoCoNuT} code~\cite{dimmelmeier05}, which solves the relativistic hydrodynamics equations, coupled to the Einstein equations for the gravitational field, within the so-called \emph{conformal flatness condition} (CFC). In the multi-dimensional case, CFC is an approximation to general relativity where the 3-metric $\gamma_{ij}$~(\ref{e:def_gij}) is conformally flat:
\begin{equation}
  \label{e:def_CFC}
  \gamma_{ij} = \psi^4 f_{ij}, 
\end{equation}
where $\psi$ is the conformal factor and $f_{ij}$ a flat 3-metric. In our particular case of spherical symmetry (no rotation), this is not an approximation but reduces to the choice of isotropic coordinates. Therefore, the whole non-rotating simulation can be exactly described within CFC, even after the formation of the black hole's apparent horizon.

Initial models of spherical neutron stars are equilibrium configurations on the unstable branch, computed in isotropic gauge using the \texttt{LORENE/rotstar\_dirac} code~\cite{lorene,lin06}. The equation of state used for generating initial data and for computing the collapse is a polytropic one, neglecting any temperature effects; pressure $p$ is related to baryon density $\bar{n}$ through
\begin{equation}
  \label{e:polytrope}
  p = \kappa \bar{n}^{\bar{\gamma}},
\end{equation}
with the adiabatic index $\bar{\gamma}=2$ and $\kappa = 4.01 \times 10^{-56}$ in SI units.

The neutron star used as initial data for the \texttt{CoCoNuT} code is a $1.62 M_\odot$ (gravitational mass) configuration, corresponding to a $1.77 M_\odot$ baryon mass. Its central density is $1.56 \, \textrm{fm}^{-3}$, central lapse $N_c = 0.40$ and circumferential equatorial radius $R_\textrm{circ} = 10.4$ km. The global accuracy indicator~\cite{bonazzola94} gives $4\times 10^{-9}$. To this configuration, we add a perturbation to the density profile ensuring that the unstable star collapses to a black hole and does not evolve to the stable branch (migration). The density profile is modified according to
\begin{equation}
  \label{e:pert_rho}
  \rho \to \rho \left[ 1 + A \sin\left( \frac{\pi r}{R_0} \right) \right],
\end{equation}
where $r$ is the coordinate radius, $A=0.01$ is the relative amplitude of the perturbation and $R_0 = 10$ km its typical size. For the \texttt{CoCoNuT} code, we use 500 radial cells on a uniform grid.

 The collapse proceeds as expected (see e.g.~\cite{cordero09}), until the formation of an apparent horizon detected by the finder described in~\cite{lin07}, at $t=0.438 \textrm{ ms}$ after the beginning of the collapse. The simulation is stopped at $t=0.495 \textrm{ ms}$, when all the matter has entered the black hole (up to numerical accuracy). The run is stopped because of the too strong increase of the gradients in many quantities (e.g. the conformal factor $\psi$) near the center of the star. This is due to the use of maximal slicing gauge condition (trace of $K_{ij}$ set to zero), which has the well-known property of
 yielding a singularity-avoiding time slicing. Nevertheless, as stated before, most of the matter has entered the black hole at that time and there is no longer evolution of the black hole. During the computation of the collapse, quantities which are used to integrate the system~(\ref{e:systXV})
 are exported from \texttt{CoCoNuT} to \texttt{GYOTO} following the procedure described in Section~3.3 of \cite{vincent11}. These quantities are the 3+1 metric and related fields $\left(N, \beta^i, \gamma_{ij}, K_{ij} \right)$, together with the fluid 4-velocity $u^\mu_{\rm fluid}$, the radius of the neutron star and the location of the black hole apparent horizon.

When integrating a null geodesic, \texttt{GYOTO} uses an interpolation at third order in the time coordinate to determine the value of the 3+1 fields at each integration step. Each geodesic is integrated backward in time until it either reaches the star's surface, or the black hole's event horizon. The difficulty here is that the location of the event horizon is not known by \texttt{CoCoNuT}, only that of the apparent horizon (that lies inside the event horizon) is known. The integration of geodesics that reach the star after the event horizon radius has become larger than the star's radius is thus non trivial. There are thus two stop conditions for geodesics that reach the central object (i.e. for geodesics not escaping towards infinity):
\begin{itemize}
\item the star's surface is hit. In such cases, the specific intensity emitted by the hit point can be computed. 
\item the fourth order Runge-Kutta adaptive step becomes smaller than a given lower limit (fixed to $10^{-6}\,M$). This latter case corresponds to a geodesic "accumulating" near the event horizon. Let us remind the reader that a backward integrated geodesic can never cross the event horizon (by the very definition of an event horizon).
\end{itemize}
The value of the integration step lower limit is chosen in such a way that the image of the event horizon on the observer's screen is smooth and spherically symmetric (choosing a too big limit results in a non spherically symmetric, noisy event horizon image).


Moreover, the norm of the 4-momentum is not conserved in the very last integration steps close to the apparent horizon. This is due to the fact that the quantity $\D t/\D\lambda$ becomes very large (it should, actually, diverge if the integration were perfect) at the event horizon: as the norm is proportional to this quantity, it is no longer conserved. As a check of this fact, the evolution of the norm divided by $\D t/\D\lambda$ has been considered. This new quantity stays close to zero even at the horizon.

\begin{figure}
\centering
	\includegraphics[width=0.225\textwidth]{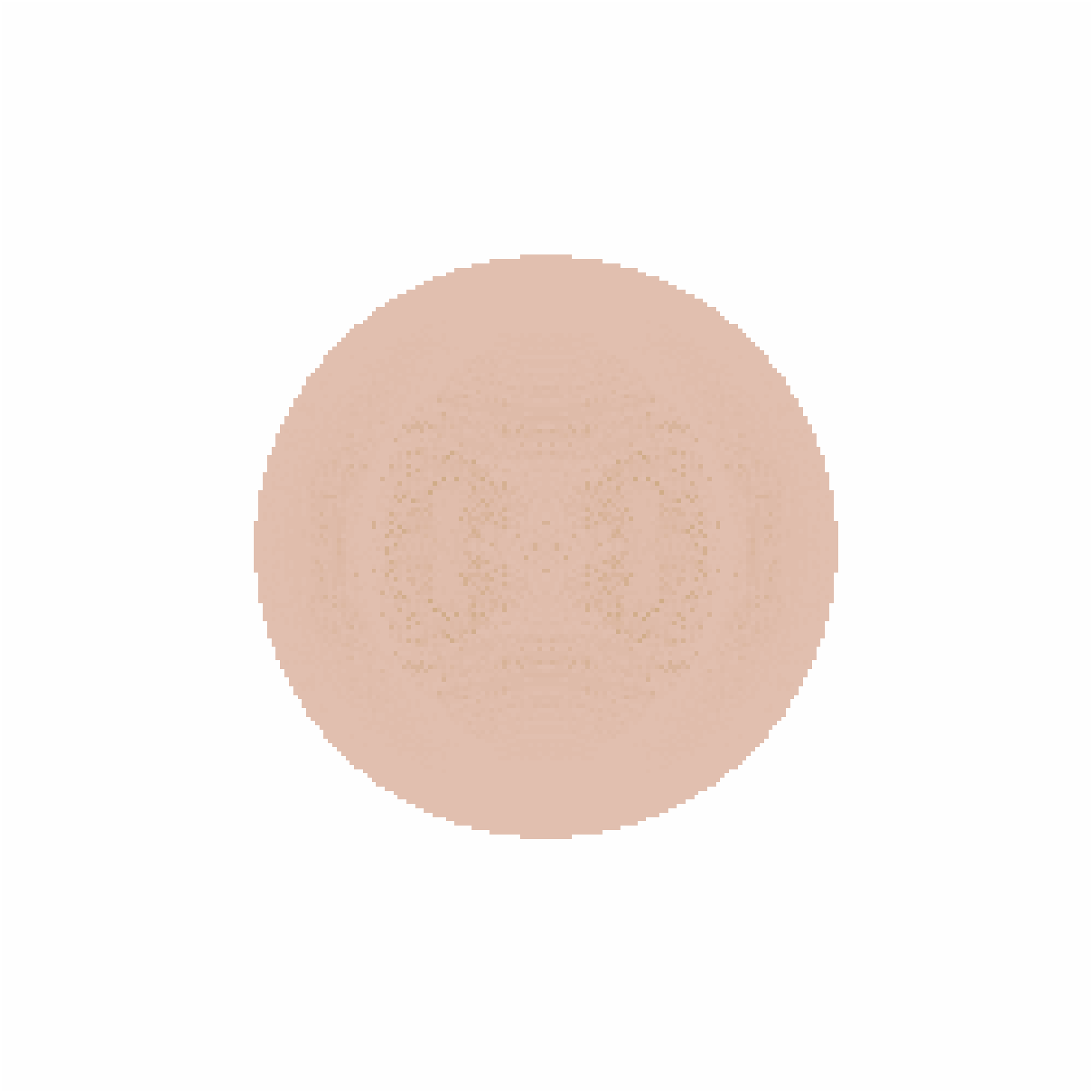}
	\includegraphics[width=0.225\textwidth]{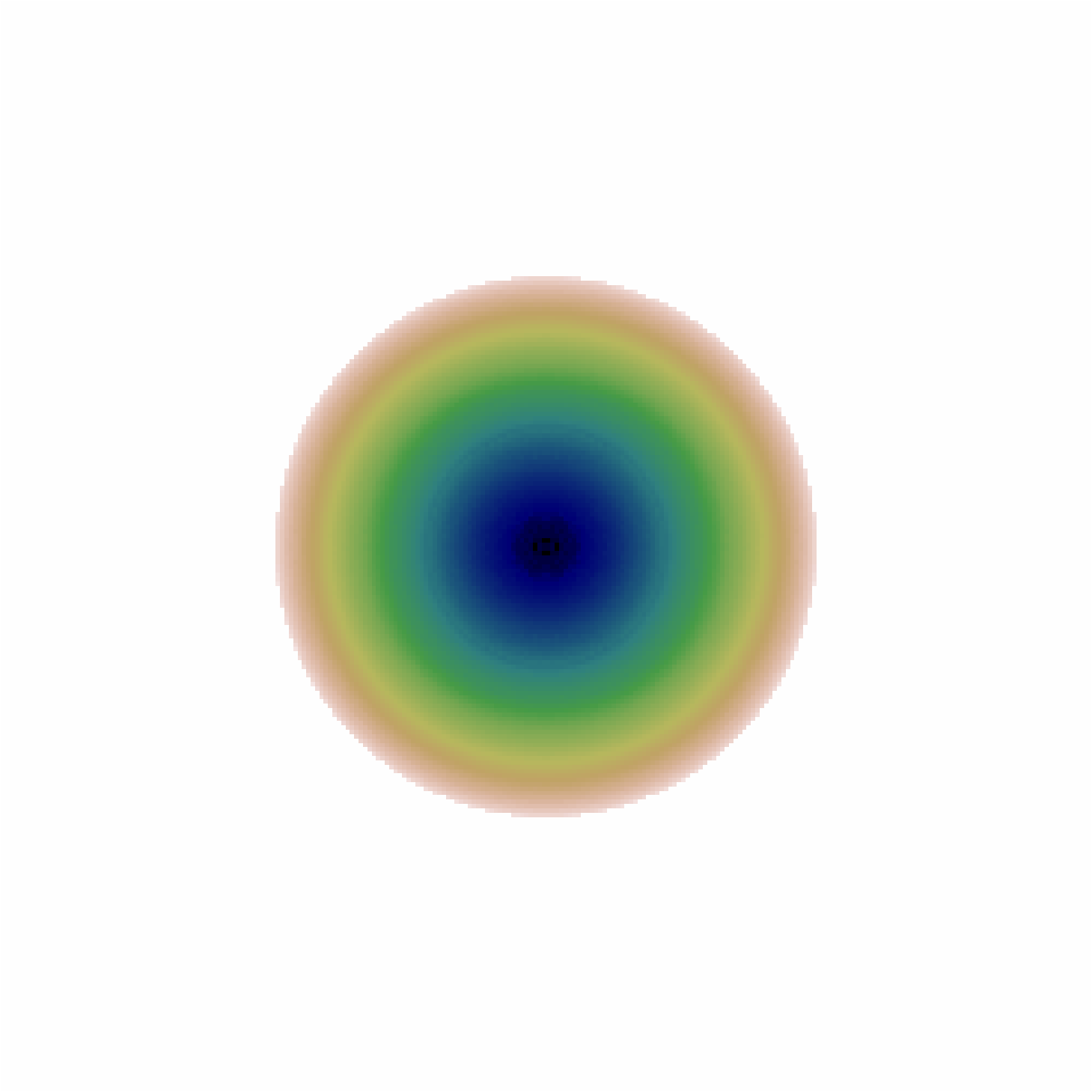}
	\includegraphics[width=0.225\textwidth]{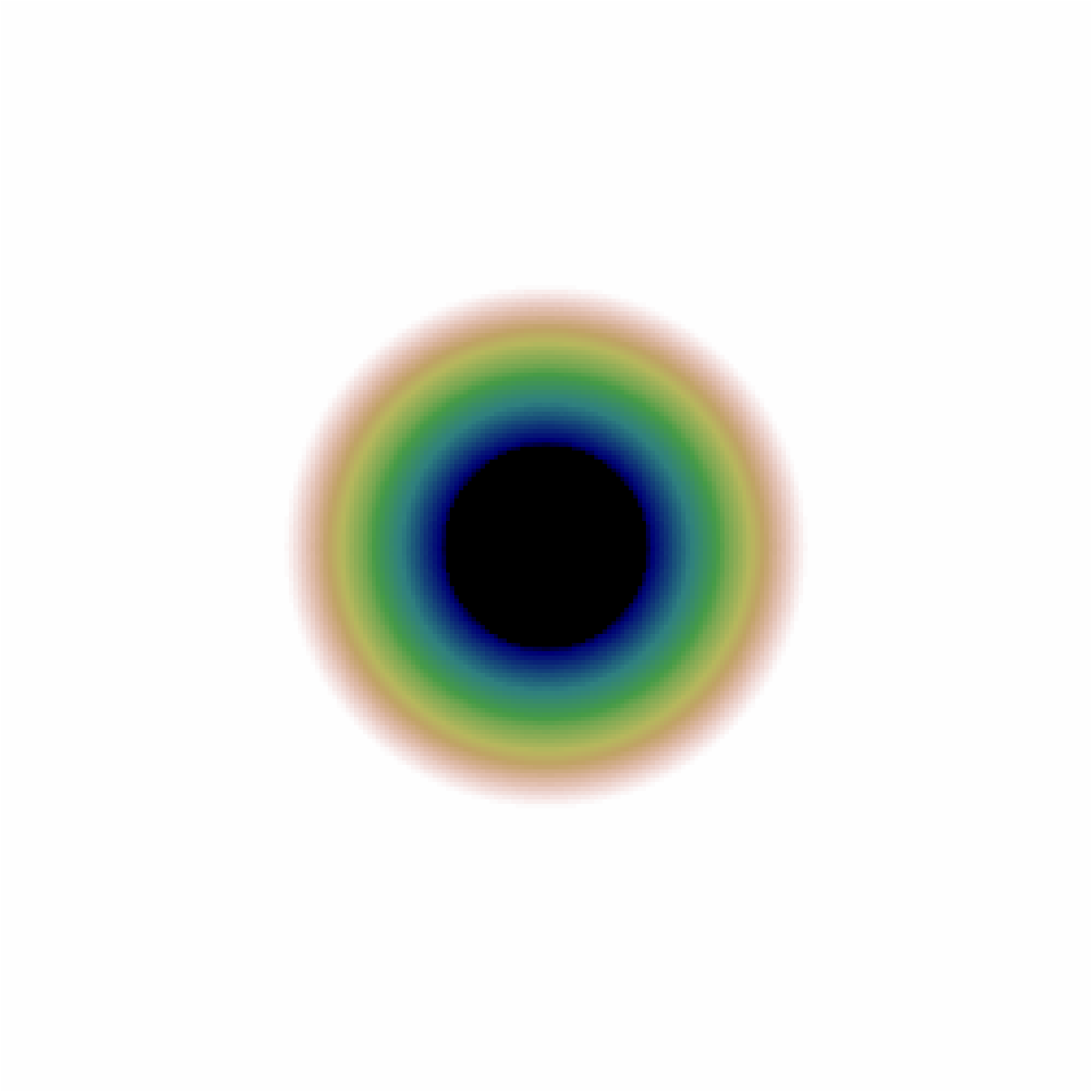}
	\includegraphics[width=0.225\textwidth]{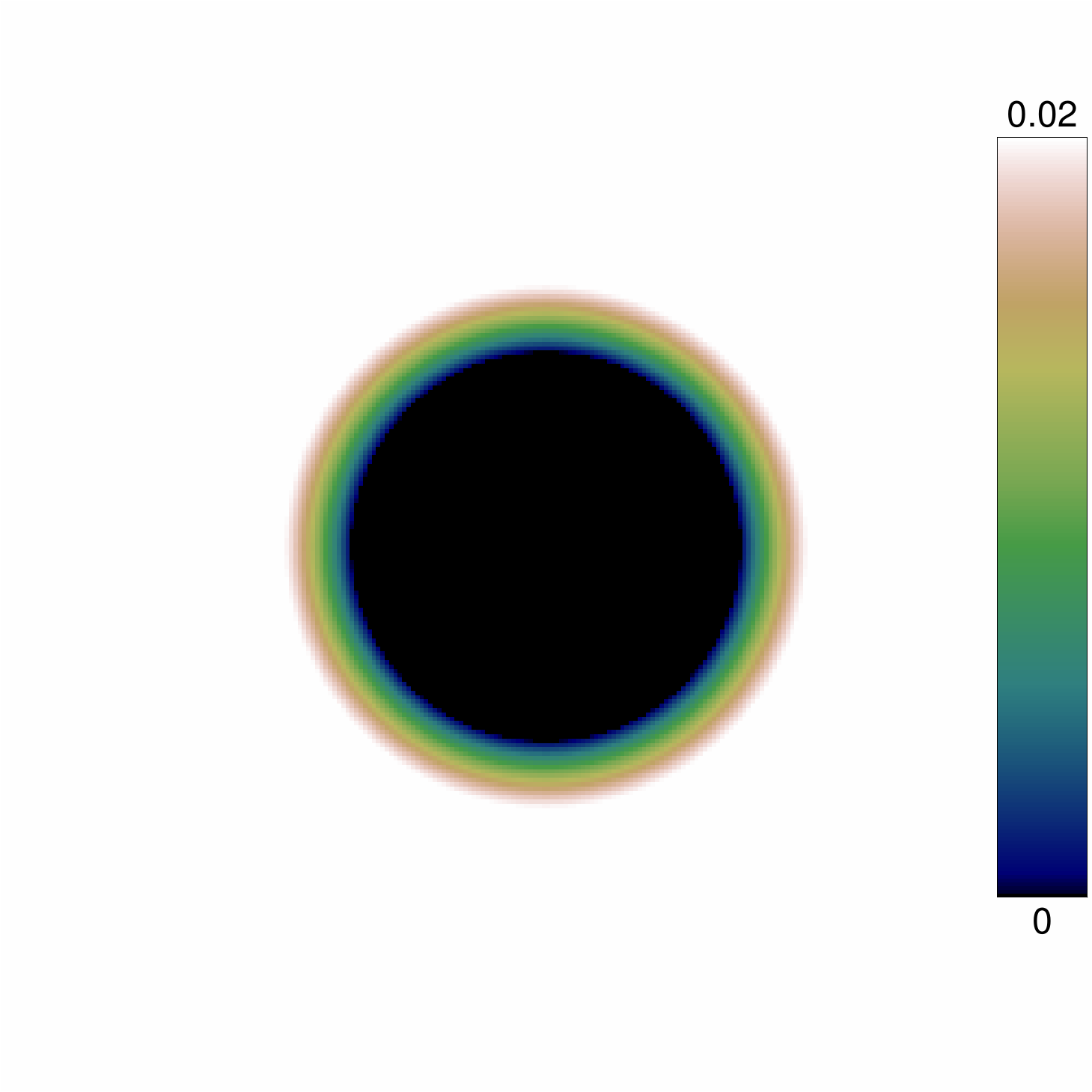}
	\caption{Images (i.e. map of specific intensity) of a non-rotating collapsing neutron star, with an optically thick surface emitting black body radiation at $10^{6}$~K. The color bar is common to the four panels and is given in SI units, $\mathrm{W\,m^{-2}\,ster^{-1}\,Hz^{-1}}$. The frequency of the photons in the observer's frame is chosen to be $10^{17}$~Hz, close to the maximum of the Planck function at $10^{6}$~K.}
	\label{f:dynaNS}
\end{figure}

Figure~\ref{f:dynaNS} shows four images of a collapsing non-rotating neutron star, as perceived by a distant observer. The first image is computed before the start of the collapse: it is thus the image of a stationary (unstable) neutron star. The other three images show different stages of the collapse, until the whole star nearly disappears below the event horizon. The intensity is shown in logarithmic scale, since the very high gravitational redshift leads to a very high dynamic range in the last images. The event horizon first appears at the center of the star due to the fact that this part of the star is closer to the observer: photons at different parts of the image have been emitted at different coordinate times $t$, later times for the central parts of the image, earlier times for the external parts of the image. As a consequence of this fact, the coordinate radius of the star at which a given photon is emitted is not the same for all pixels in a given image: it is shorter at the center of the image (more evolved part) than at the edge. For instance, the coordinate radius of the star at the emission of the photon reaching the central pixel of the second image on the left is of $2.7$~km while it is $4.7$~km on a pixel located at the edge of the star's image. The coordinate radius of the star on the left image is of about $7$~km (here, the coordinate radius is the same at any pixel on the image, as the left star is stationary, and smaller than the circumferential radius cited above of $10.4$~km), whereas the coordinate radius of the star in the rightmost image is of approximately $2.9$~km for a pixel at the edge of the image. The fact that the ratio of the apparent radii of the star on the left and right panels of Figure~\ref{f:dynaNS} is much less than $7/2.9\simeq 2.4$ is due to the strong bending of geodesics in the vicinity of the nascent black hole, resulting in its apparent radius being larger as the object becomes more compact. 
This effect is exactly the same as the one which makes the black hole shadow (the black area in the image 
of a black hole in front of an emitting region) larger than the projected size of the event horizon.
For a Schwarzschild black hole, the enlarging factor is $3\sqrt{3} / 2 \simeq 2.6$. 
This explains why the shrink of the size of the image of the collapsing star shown in Figure~\ref{f:dynaNS} is not very pronounced. 
Finally, let us stress the fact that the time elapsed between the appearance of the event horizon and the disappearance of the whole star behind it is extremely short. It is of approximately $0.2$~ms in the observer's frame.

Figures~\ref{f:statNS} and~\ref{f:dynaNS} are the first examples of ray-traced images in numerically computed spacetimes. In the present work, the physics of emission at the surface of the neutron star has not been studied in detail at all. In particular, the emission during the collapse will most certainly be quite different from a simple blackbody. However the aim of the computations presented here is not (yet) to propose astrophysically relevant images, but to give first examples of the interest of such a ray-tracing code as \texttt{GYOTO}, capable of integrating geodesics in numerical spacetimes. Future works will be devoted to applying this code to diverse astrophysically realistic scenarios.


\section{Conclusion}
\label{sec:conc}

We have re-expressed the geodesic equation within the framework of the 3+1 formalism of general relativity,
obtaining equations (\ref{e:dEdt}), (\ref{e:systXV}) and (\ref{e:redshift}).
Equation~(\ref{e:dEdt}), ruling the evolution of the particle energy with respect to Eulerian observers, 
has already been derived (in an equivalent form) by Merl{\'\i}n and Salgado \cite{merlin11}. On the other hand, the system (\ref{e:systXV}) for the position of the geodesic and the redshift formula (\ref{e:redshift}) are novel. In particular, (\ref{e:systXV}) significantly differs from previous 3+1 geodesic equations in  the literature \cite{HugheKWWST94}, as discussed in Section~\ref{s:3p1_geod_coord}.
The 3+1 equations have been implemented in the ray-tracing code \texttt{GYOTO} \cite{vincent11,gyoto},
which has enabled us to compute images of stationary and collapsing neutron star numerical spacetimes generated
by the \texttt{LORENE/nrotstar} \cite{gourgoulhon10b,lorene} and \texttt{CoCoNuT}~\cite{dimmelmeier05} codes.

Future work will be devoted to the development of ray-tracing computations in numerical spacetimes for astrophysically relevant problems. 
In particular, we shall try to carry on computations of images and spectra of astrophysical phenomena in the vicinity of compact objects, which can be alternative to black holes.
This work is of particular interest in the perspective of a direct test of the nature of the central compact object of the Galaxy, Sgr~A*~(see the review \cite{genzel10}, 
and in particular the section devoted to the alternatives to the black hole case).


The capability of \texttt{GYOTO} to integrate geodesics in numerical spacetimes will be very interesting too, in order to visualize spacetimes, be it binary
black holes spacetimes, binary neutron stars spacetimes, black hole -- neutron star binary spacetimes, or any other interesting
metric (see reviews on these topics by~\cite{hinder10,mcwilliams11,shibata11,faber12}) . \texttt{GYOTO} could be used to image a background sky of stars, or a simple coordinate grid,
putting in light the effect of strong gravity on these background objects.


\ack
The authors thank the anonymous referees for interesting comments that allowed to improve the quality of the article.
This work was supported by grants from R\'egion Ile-de-France and by the ANR grant 06-2-134423 \emph{M\'ethodes math\'ematiques pour la relativit\'e g\'en\'erale}.


\appendix

\section{Derivation for a single geodesic}  \label{s:derive_single}

Here we consider a single (null or timelike) geodesic $\LL$ and not a congruence as in 
Sect.~\ref{s:geod_congruence}. In the context of the 3+1 formalism, a natural 
parameter along $\LL$ is the time coordinate $t$ that labels the  foliation $(\Sigma_t)_{t\in\mathbb{R}}$. 
A priori, $t$ is not an affine parameter along $\LL$; it is related to 
the affine parameter associated to $p^\alpha$ by (\ref{e:dlambdadt}). The vector tangent to $\LL$
associated to the parametrization by $t$ is $q^\alpha := \D x^\alpha / \D t$, $\D x^\alpha$ being the infinitesimal displacement vector along $\LL$ corresponding to the infinitesimal 
parameter increment $\D t$. The vector $q^\alpha$ obeys $q^\mu \nabla_\mu t = 1$. Moreover, $q^\alpha$ and $p^\alpha$ being vectors both tangent to $\LL$, they must be collinear. We deduce then from (\ref{e:decomp_p})
that 
\be \label{e:decom_q}
  q^\alpha = N (n^\alpha + V^\alpha) . 
\ee
Here, the fact that the proportionality coefficient in (\ref{e:decom_q}) is $N$ comes from 
$q^\mu \nabla_\mu t = 1$, $n^\mu \nabla_\mu t = N^{-1}$ and $V^\mu \nabla_\mu t = 0$. 
Taking into account (\ref{e:decomp_p}), the geodesic equation (\ref{e:p_geod}) is equivalent to 
\be
  q^\mu \nabla_\mu \left[ E (n^\alpha + V^\alpha) \right] = 0 . 
\ee
Expanding and using (\ref{e:nab_n}), as well as (\ref{e:decom_q}), we get 
\be \label{e:geod_q_nab}
  q^\mu \nabla_\mu  E \;  (n^\alpha + V^\alpha)
  + E\left( D^\alpha N - N K^\alpha_{\ \, \mu} V^\mu + q^\mu \nabla_\mu V^\alpha \right) = 0 . 
\ee
Note that the only derivatives of $E$ and $V^\alpha$ that appear in this equation are derivatives along $\LL$ (through
the operator $q^\mu\nabla_\mu$). Consequently (\ref{e:geod_q_nab}) is valid for a single geodesic, contrary to
(\ref{e:geod_3p1_prov2}) which holds only for a congruence. 

The projection of (\ref{e:geod_q_nab}) along $n^\alpha$ yields
\[
  - q^\mu \nabla_\mu E + E q^\mu n_\nu \nabla_\mu V^\nu = 0 ,
\]
where we have used $n_\nu V^\nu = 0$, $n_\nu D^\nu \ln N = 0$ and $n_\nu K^\nu_{\ \, \mu} = 0$. 
Now, since $n_\nu V^\nu = 0$, we have $n_\nu \nabla_\mu V^\nu = - V^\nu \nabla_\mu n_\nu$, i.e.,
thanks to (\ref{e:nab_n}), $n_\nu \nabla_\mu V^\nu = V^\nu (K_{\mu\nu} + D_\nu \ln N \, n_\mu)$. 
In addition, from the very definition of the vector $q^\alpha$, $q^\mu \nabla_\mu E = \D E/\D t$. 
We thus end with 
\be \label{e:dEdt_single}
  \frac{\D E}{\D t} = E V^j \left( N K_{jk} V^k - \partial_j N \right) , 
\ee
which is exactly (\ref{e:dEdt}). 

The orthogonal projection of (\ref{e:geod_q_nab}) onto $\Sigma_t$ is performed via the operator
$\gamma^\alpha_{\ \, \nu}$. Thanks to the properties
$\gamma^\alpha_{\ \, \nu} n^\nu = 0$, $\gamma^\alpha_{\ \, \nu} V^\nu = V^\alpha$, 
$\gamma^\alpha_{\ \, \nu} D^\nu N = D^\alpha N$ and $\gamma^\alpha_{\ \, \nu} K^\nu_{\ \, \mu} = 
K^\alpha_{\ \, \mu}$, this yields
\be \label{e:qnabV_prov}
  q^\mu \nabla_\mu E \; V^\alpha
  + E \left( D^\alpha N - N K^\alpha_{\ \, \mu} V^\mu + \gamma^\alpha_{\ \, \nu} q^\mu \nabla_\mu V^\nu \right) = 0 . 
\ee
To evaluate the last term, let us expand the velocity vector $V^\alpha$ onto the coordinate basis
$(\eb_\alpha) := (\partial/\partial x^\alpha)$ associated to the coordinates $(x^\alpha)=(t,x^i)$. Since $V^\alpha$ is tangent to $\Sigma_t$ it has no
component along $\eb_0 = \partial/\partial t$: 
\be
  V^\alpha = V^j \, \eb_j^\alpha . 
\ee
We then have, since $\gamma^\alpha_{\ \, \nu}  \eb_j^\nu = \eb_j^\alpha$ (for $\eb_j$
is tangent to $\Sigma_t$), 
\begin{eqnarray}
  \gamma^\alpha_{\ \, \nu} q^\mu \nabla_\mu V^\nu & = & q^\mu \nabla_\mu V^j \; \eb_j^\alpha
  + V^j \gamma^\alpha_{\ \, \nu} q^\mu \nabla_\mu \eb_j^\nu  \nonumber \\
  & = & q^\mu \nabla_\mu V^j \; \eb_j^\alpha
  + N V^j \gamma^\alpha_{\ \, \nu} (n^\mu + V^\mu)  \nabla_\mu \eb_j^\nu  \nonumber \\
 & = & q^\mu \nabla_\mu V^j \; \eb_j^\alpha
  + N V^j \left( \gamma^\alpha_{\ \, \nu} n^\mu \nabla_\mu \eb_j^\nu 
  + V^\mu D_\mu \eb_j^\alpha \right) . \label{e:pqnab_V}
\end{eqnarray}
Now, thanks to (\ref{e:n_nab_V}) with $V^\alpha$ replaced by $\eb_j^\alpha$, we can write 
\begin{eqnarray}
  \gamma^\alpha_{\ \, \nu} n^\mu \nabla_\mu \eb_j^\nu  & = & 
  \gamma^\alpha_{\ \, \nu} \left( N^{-1} \Lie{m} \eb_j^\nu - K^\nu_{\ \, \mu} \eb_j^\mu
  + \eb_j^\mu D_\mu \ln N \, n^\nu \right) \nonumber \\
 & = & \gamma^\alpha_{\ \, \nu} N^{-1} \left(\Lie{\eb_0} \eb_j^\nu 
  - \Lie{\beta} \eb_j^\nu \right ) - K^\alpha_{\ \, \mu} \eb_j^\mu  \nonumber \\
 & = & N^{-1} \Lie{\eb_j} \beta^\alpha - K^\alpha_{\ \, \mu} \eb_j^\mu , \label{e:pnnab_dj}
\end{eqnarray}
where we have used (\ref{e:Lie_m_t_beta}), $\gamma^\alpha_{\ \, \nu} K^\nu_{\ \, \mu} = K^\alpha_{\ \, \mu}$, $\gamma^\alpha_{\ \, \nu} n^\nu = 0$,
$\Lie{\eb_0} \eb_j = [\eb_0,\eb_j] = 0 $
(since $\eb_0$ and $\eb_j$ are vectors of a coordinate basis), 
$\Lie{\beta} \eb_j^\alpha = - \Lie{\eb_j} \beta^\alpha$,
and $\gamma^\alpha_{\ \, \nu} \Lie{\eb_j} \beta^\nu = \Lie{\eb_j} \beta^\alpha$
(since $\eb_j^\alpha$ and $\beta^\alpha$ are both tangent to $\Sigma_t$). 
Thanks to (\ref{e:pnnab_dj}), (\ref{e:pqnab_V}) becomes
\be
  \gamma^\alpha_{\ \, \nu} q^\mu \nabla_\mu V^\nu  = 
   q^\mu \nabla_\mu V^j \; \eb_j^\alpha 
  + V^j \left[ \Lie{\eb_j} \beta^\alpha + N \left( V^\mu D_\mu \eb_j^\alpha
  - K^\alpha_{\ \, \mu} \eb_j^\mu \right) \right] . 
\ee
Substituting this expression into (\ref{e:qnabV_prov}) and setting $\alpha = i$ (for $\alpha =0$, 
the equation reduces to $0=0$), we get 
\[
  \fl q^\mu \nabla_\mu V^j \; \eb_j^i = - E^{-1} q^\mu \nabla_\mu E \; V^i 
   V^j \left[ N \left( K^i_{\ \, k} \eb_j^k - V^k D_k \eb_j^i
  \right) - \Lie{\eb_j} \beta^i \right]   - D^i N + N K^i_{\ \, j} V^j .
\]
Let us consider the components of this vector equation in the coordinate basis $(\eb_j)$. 
We then have $\eb_j^i = \delta^i_{\ \, j}$, 
$D_k \eb_j^i = {}^3\Gamma^i_{\ \, jk}$,
$D^i N = \gamma^{ij} \partial_j N$
and, by the definition 
of a Lie derivative, $\Lie{\eb_j} \beta^i = \partial_j \beta^i$. Since in addition, 
$q^\mu \nabla_\mu V^j = \D V^i / \D t$ and $q^\mu \nabla_\mu E = \D E / \D t$, we get 
\be
  \frac{\D V^i}{\D t} = -\frac{1}{E}\frac{\D E}{\D t} \, V^i
  + N V^j \left( 2 K^i_{\ \, j}  - {}^3 \Gamma^i_{\ \, jk} V^k \right)
  - \gamma^{ij} \partial_j N - V^j \partial_j \beta^i . 
\ee
Substituting (\ref{e:dEdt_single}) for $\D E / \D t$, we recover (\ref{e:dVdt}).

\section{Second order form of the 3+1 geodesic equation} \label{s:second_order}

The standard (4-dimensional) geodesic equation is
\be \label{e:geod_4D}
	\frac{\D^2 X^\alpha}{\D \lambda^2} + {}^4 \Gamma^\alpha_{\ \, \mu\nu} \frac{\D X^\mu}{\D\lambda}
	\frac{\D X^\nu}{\D\lambda} = 0 , 
\ee 
where (i) $X^0 := t$, in addition to the $X^i$'s defined by (\ref{e:X_t}), 
(ii) the ${}^4 \Gamma^\alpha_{\ \, \mu\nu}$'s are 
the Christoffel symbols of the spacetime metric $g_{\alpha\beta}$ and 
(iii) $\lambda$ is an affine parameter along the particle's worldline $\LL$.  
If $\LL$ is timelike, $\lambda$ is equal to a constant times the particle's proper time $\tau$. 
More specifically, if $\lambda$ is the affine parameter associated to the particle's 4-momentum
and $m$ is the particle's mass,  
we deduce from the relations $p^\alpha = \D x^\alpha / \D\lambda$ and 
$p^\alpha = m \, \D x^\alpha / \D\tau$ that $\lambda = \tau / m$.

Let us check that (\ref{e:geod_4D}) can be recovered from the system (\ref{e:systXV}). 
Extracting $V^i$ from (\ref{e:dXdt}), substituting it in (\ref{e:dVdt}) and expanding, we obtain 
a second order differential equation for $X^i(t)$: 
\bea
  \fl {\ddot X}^i + 
  \frac{1}{N} \left[ K_{jk} ( {\dot X}^j + \beta^j )
   ( {\dot X}^k + \beta^k ) - ( 2 {\dot X}^j + \beta^j ) \partial_j N  
  - \der{N}{t}  \right]
  {\dot X}^i \nonumber \\
  + 2 \left[ D_j \beta^i - N K^i_{\ \, j} + \frac{\beta^i}{N} 
   ( K_{jk} \beta^k - \partial_j N ) \right] {\dot X}^j  
  + \left( {}^3\Gamma^i_{\ \, jk} + \frac{\beta^i}{N} K_{jk} \right)
  {\dot X}^j {\dot X}^k \nonumber \\
 + N \gamma^{ij} \partial_j N - 2N K^i_{\ \, j} \beta^j   
 + \frac{\beta^i}{N} \left( K_{jk} \beta^j \beta^k  
  - \der{N}{t} - \beta^j \partial_j N \right) \nonumber \\
   + \der{\beta^i}{t} + \beta^j D_j \beta^i 
	= 0 . \label{e:geod3p1_2nd_order}
\eea

On the other hand, the ${}^4 \Gamma^\alpha_{\mu\nu}$'s appearing in (\ref{e:geod_4D})
can be expressed in terms of the 3+1 quantities as follows (cf. Appendix~B of \cite{Alcub08}): 
\bea
  \fl {}^4 \Gamma^0_{\ \, 00}= \frac{1}{N} \left( \der{N}{t} +  \beta^j \partial_j N
  - K_{jk} \beta^j \beta^k \right) \label{e:4Gam_000} \\
  \fl  {}^4 \Gamma^0_{\ \, 0j} = \frac{1}{N} \left( \partial_j N - K_{jk} \beta^k \right) \\
  \fl {}^4 \Gamma^0_{\ \, jk} = - \frac{1}{N} K_{jk} \label{e:4Gam_0jk} \\
  \fl {}^4 \Gamma^i_{\ \, 00} =	N \gamma^{ij} \partial_j N 
  - 2N K^i_{\ \, j} \beta^j + \frac{\beta^i}{N} \left( K_{jk} \beta^j \beta^k - \der{N}{t} -  \beta^j \partial_j N
  \right)  + \der{\beta^i}{t} + \beta^j D_j \beta^i \label{e:4Gam_i00}\\ 
 \fl  {}^4 \Gamma^i_{\ \, 0j} = D_j \beta^i  - N K^i_{\ \,j} 
  + \frac{\beta^i}{N} \left( K_{jk} \beta^k  - \partial_j N \right) \\
  \fl {}^4 \Gamma^i_{\ \, jk} = {}^3 \Gamma^i_{\ \, jk} + \frac{\beta^i}{N} K_{jk} . \label{e:4Gam_ijk}
\eea 
In addition, we have
\bea
  \frac{\D X^\alpha}{\D \lambda} = {\dot X}^\alpha \frac{\D t}{\D \lambda} 
  \label{e:dXdlamb} \\
  \frac{\D^2 X^\alpha}{\D \lambda^2} = {\ddot X}^\alpha 
  \left( \frac{\D t}{\D \lambda} \right)^2 + {\dot X}^\alpha \frac{\D^2 t}{\D \lambda^2} . \label{e:d2Xdlamb2}
\eea
Accordingly, for $\alpha=0$, (\ref{e:geod_4D}) becomes 
(note that ${\dot X}^0 = 1$ and ${\ddot X}^0 = 0$):
\[
  \frac{\D^2 t}{\D \lambda^2} + 
  \left( \frac{\D t}{\D \lambda} \right) ^2 
  \left( {}^4 \Gamma^0_{\ \, 00} + 2 \, {}^4 \Gamma^0_{\ \, 0j} {\dot X}^j
  + {}^4 \Gamma^0_{\ \, jk} {\dot X}^j {\dot X}^k \right) = 0 . 
\] 
In view of (\ref{e:4Gam_000})-(\ref{e:4Gam_0jk}), we get 
\be \label{e:d2tdlamb2}
  \fl  \left( \frac{\D t}{\D \lambda} \right) ^{-2}  \frac{\D^2 t}{\D \lambda^2} = 
  \frac{1}{N} \left[ K_{jk} ({\dot X}^j + \beta^j)({\dot X}^k + \beta^k)
  - (2{\dot X}^j + \beta^j) \partial_j N - \der{N}{t} \right] . 
\ee 
For $\alpha=i$, (\ref{e:geod_4D}) becomes,
thanks to (\ref{e:dXdlamb})-(\ref{e:d2Xdlamb2}), 
\[
  {\ddot X}^i + \left( \frac{\D t}{\D \lambda} \right) ^{-2}  \frac{\D^2 t}{\D \lambda^2} \, {\dot X}^i + {}^4 \Gamma^i_{\ \, 00}
   + 2\, {}^4 \Gamma^i_{\ \, 0j}{\dot X}^j 
  + {}^4  \Gamma^i_{\ \, jk} {\dot X}^j {\dot X}^k = 0 . 
\]
In view of (\ref{e:d2tdlamb2}) and (\ref{e:4Gam_i00})-(\ref{e:4Gam_ijk}), 
we recover (\ref{e:geod3p1_2nd_order}).

\end{document}